\documentclass[12pt]{article}
\usepackage{amssymb,amsmath,epsfig}

\def\be{\begin{equation}}
\def\ee{\end{equation}}
\def\bea{\begin{eqnarray}}
\def\eea{\end{eqnarray}}
\def\ba{\begin{array}{rcl}}
\def\ea{\end{array}}
\def\der{\partial}
\def\tfract#1/#2{{\textstyle{\raise0.8pt\hbox{$\scriptstyle#1$}\over%
\hbox{\lower0.8pt\hbox{$\scriptstyle#2$}}}}}
\def\mezzo{\tfract 1/2 }
\def\trequarti{\tfract 3/4}

\def\quarto{\tfract 1/4}

\def\downnormalfill{$\,\,\vrule depth4pt width0.4pt
\leaders\vrule depth 0pt height0.4pt\hfill\vrule depth4pt width0.4pt\,\,$}
\def\WT#1{\mathop{\vbox{\ialign{##\crcr\noalign{\kern3pt}
      \downnormalfill\crcr\noalign{\kern1.8pt\nointerlineskip}
      $\hfil\displaystyle{#1}\hfil$\crcr}}}\limits}

\catcode`@=11 
\font\titlefontB=cmssdc10 at 20pt
\catcode`@=12 
\fussy
\begin{document}


\vskip 1.7 truecm 

\centerline {\titlefontB Gravitons scattering from classical matter}

\vskip 3.2 truecm

\centerline  {\bf E. GUADAGNINI}

\vskip 1 truecm 

\centerline {Dipartimento di Fisica {\sl Enrico Fermi} dell'Universit\`a di Pisa,}
\centerline {and INFN Sezione di Pisa, Italy} 

\vskip 4 truecm 

\noindent {\bf Abstract.}  ~The low energy scattering of gravitons from a composite extended system, which is made of classical massive bodies,  is considered; by using the Feynman rules of effective quantum gravity, the corresponding cross-section is computed to lowest order in powers of the gravitational coupling constant. For the gravitons scattering from a rotating planet or a star, it is shown that the classical limit of the matter-gravitons coupling in the effective quantum gravity lagrangian  leads to a low energy scattering amplitude which coincides with the expression obtained in classical general relativity. 
 
\vskip 1.4 truecm 

 

\vskip 1.4 truecm

\noindent {\hrule height0.2pt width200pt depth0pt}

\medskip 

\medskip

\noindent E-Mail:   ~enore.guadagnini@df.unipi.it 

\hfill

\vfill\eject

\section{Introduction}

In the low energy limit, the  scattering of electromagnetic waves from free charged particles can be approximated by the Thomson scattering, in which  the outcoming radiation can be interpreted as  the  radiation  emitted because of the particles acceleration which is induced by the incoming wave.  The resulting total cross-section is given by the Thomson formula  $\sigma = (8 \pi/3 ) \,  r_{\rm cl}^2  $,  where $ r_{\rm cl} = q^2 / m c^2$ represents the classical radius of the charged particles. 
Instead,  the gravitons scattering from classical free massive bodies is expected to be dominated ---in the same limit---  by the newtonian scattering, which is due to the gravitational attraction between gravitons and massive bodies.  When the interaction  potential  vanishes at large distances as the inverse power of the distance, the total cross section is divergent.  The low energy dominance of the newtonian scattering  is in agreement with all the results ---which have been obtained by means of classical arguments--- concerning, for instance,  the graviton scattering from black holes \cite{RB, FHM, SC, FN, AJ} and the low energy  gravitons scattering from a  planet or a star \cite{P, S, W, MR, DMB, DL, BE, D1, D2}. The classical arguments are essentially based on the study of the linearized gravitational equations in a nontrivial background,  possibly with the introduction of appropriate Green's functions, or by means of the JWKB or the partial waves methods. 
As far as the gravitons scattering from elementary particles is concerned,    the computation of the cross-section  for the  scattering of gravitons from scalar or Dirac particles, for examples, can be  found in the references  \cite{VL, C, DW, GJ, B, AV, BG, CN}.    

One of the purposes of this article is to produce the expression of the cross-section for the gravitons scattering from a composite extended system which is made of classical free massive bodies.  The effective quantum field theory description  of gravity will be used to derive  the corresponding scattering amplitude. It will be shown that, in the appropriate semiclassical  limit, the cross-section for the graviton scattering from a rotating star is also recovered; the result agrees with classical Peters formula \cite{P} and agrees with the expression obtained in  classical  general relativity. 

In effective quantum gravity, the newtonian contribution to the Compton graviton-scalar amplitude is described, in the Born approximation,  by a single Feynman diagram containing one 3-gravitons  vertex and one graviton propagator in the $t$-channel. This single diagram (with the appropriate modifications in the external legs) is expected to give the dominant part of the the low energy scattering amplitude also in the case of gravitons scattering from a classical massive body like a planet or a star. More precisely,  the classical limit of the matter-gravitons coupling in the effective quantum gravity lagrangian  leads to a scattering amplitude which corresponds to a single diagram containing one 3-gravitons  vertex. However,  the computation of this Feynman diagram which has been presented in \cite{K}   is not in agreement with Peters formula \cite{P};  it is not in agreement also with the results  obtained in classical general relativity and doubts   \cite{K} on the gauge-invariance of the result  have been raised. The second purpose  of this article is to clarify this issue and to show that, really,  the contribution of the 3-gravitons Feynman diagram (with modified external legs)  to the  transition amplitude is in complete agreement with Peters equation  \cite{P}, it  is in complete agreement with the results which have been obtained by means of classical arguments and represents the low energy approximation of a  gauge-invariant expression. This subject presents some interest because, since the one-graviton exchange process contains one 3-gravitons vertex,  the gravitons scattering from classical matter at low energy provides a test of  the non-linear structure of the equations  which describe the dynamics of the gravitational field in general relativity. 

This article is organized as follows. 
In Section~2, the scattering of gravitons from a classical matter system made of a set of dust particles is considered and the corresponding cross-section is computed, at first order in powers of the gravitational coupling constant, by means of the effective quantum field theory formalism.  The gauge-invariant  transition amplitude of the process is written as a linear combination of the amplitudes which refer to the gravitational scattering from the elementary constituents of the dust, which are approximated by spinless massive particles. The low energy behaviour of the the amplitude  is considered and its expression  is produced  as a function of the  velocities of the massive bodies.  The corrections to the geometric optics approximation are computed at first order in powers of the graviton momentum transfer and at first order in powers of the velocities of the massive bodies; the corresponding transition amplitude only depends on the total energy and total angular momentum of the matter system and  coincides with the amplitude for the graviton scattering from a rotating planet or star. In Section~3 it is shown that the same result can also be obtained by considering,  in the quantum field theory approach, the classical limit for the lagrangian matter-gravitons  coupling; in this case, the transition amplitude is given by a single Feynman diagram with precisely one 3-gravitons vertex and the result  coincides with the expression obtained in classical general relativity. 
  
\section{Cross-section for the gravitons scattering} 

\begin{figure}[htbp]
\centerline{\includegraphics[width=2.70in]{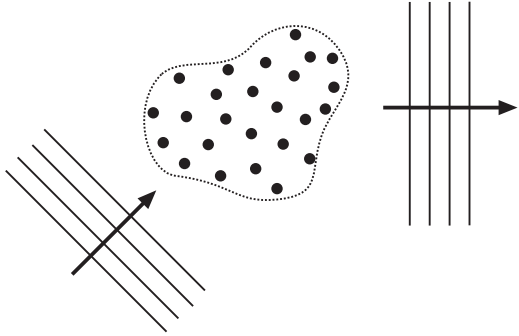}}
\caption{Gravitons scattering from classical matter.}
\label{F1}
\end{figure}

Let us consider the gravitons scattering from a classical matter system made of dust particles; as depicted in Figure~\ref{F1},  this system can be represented by a collection of free moving classical massive objects. It is  assumed that the motion of each elementary constituent of the dust is influenced only by the gravitational field.  It is also assumed that the typical size $L$ of each particle is  small compared to the characteristic wave length $\lambda $ of the gravitational wave, $L/\lambda \ll 1$, so that, as far as the graviton scattering is concerned, each constituent of the dust can be approximated by a pointlike particle. Since we are interested in low frequence gravitational waves, depending on the value of $\lambda$ the massive particles system could consist of a  Boltzmann molecular gas, or it could be made by a large number of massive bodies with the size varying from the micron scale up to o few kilometers or, possibly,  up to the planet length scale. 

\vskip 0.1 truecm

The gravitational scattering amplitude  can be written as a sum of amplitudes  for the scattering of one graviton from a single classical particle of dust. The expression of the amplitude for the elementary graviton-particle scattering can be obtained by taking the low-energy limit  of  the so-called Compton  graviton-scalar amplitude computed in the Born approximation.

\subsection{Single particle scattering}

\begin{figure}[htbp]
\centerline{\includegraphics[width=4.30in]{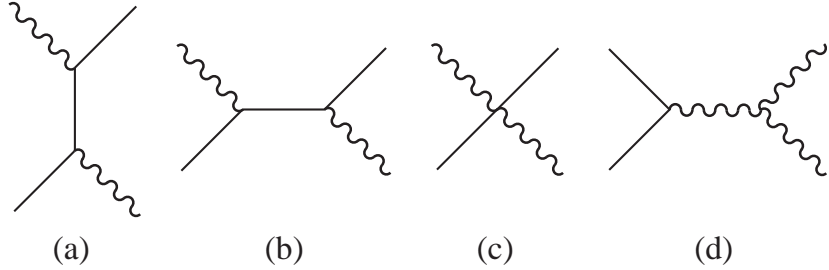}}
\caption{Feynman diagrams entering the Compton graviton-scalar scattering.}
\label{F2}
\end{figure}

The coupling of gravitons with massive scalar particles is described by the action $S$ for the gravitational and matter fields;  $S$   is the sum  of the action of a minimally coupled massive scalar field and the Einstein-Hilbert action of general relativity
\be
S = \int d^4x \, \sqrt{ - g\, }  \,  \left \{   \mezzo \left [ g^{\mu \nu } \der_\mu \varphi \, \der_\nu \varphi - m^2 \varphi^2  \right ] -   ( 16 \pi G)^{-1}  \, R (x) \right \} \; . 
\label{1}
\ee 
One can put  $ g_{\mu \nu} (x) = \eta_{\mu \nu} + h_{\mu \nu}(x) $,  where $\eta_{\mu \nu} $ denotes the Minkowski flat metric and $h_{\mu \nu}(x)$ represents the small fluctuation of the metric. The expansion of the functional (\ref{1}) in powers of $h_{\mu \nu }$ provides the interaction vertices which can be used to compute the amplitude for the graviton scattering \cite{G, F, RF}.  The term of this expansion which is quadratic  in $h_{\mu \nu }$  together with the gauge-fixing  lagrangian determine the form of the graviton propagator. 
To lowest order in powers of the gravitational coupling constant $G$, the Compton amplitude $A_C$ is given by the sum of the contributions which are associated with the Feyman diagrams  shown in Figure~\ref{F2} 
\be 
A_C = A_{(a)} + A_{(b)} + A_{(c)} + A_{(d)} \; . 
\label{2}
\ee
For generic values of the particle  momenta, in order to recover a gauge-invariant amplitude one has to sum the contributions of all the diagrams shown in Figure~2. 

\vskip 0.2 truecm

\vskip 0.2 truecm 

 In what follows, we shall use the standard units in which $\hbar = c = 1$. Let us denote by $p_1 = ( E_1 , \vec p_1 )$ and $p_2= (E_2 , \vec p_2 )$ the initial and final momenta of the massive scalar particle;  the incoming graviton has momentum $k_1= ( \omega_1 , \vec k_1 ) $ and polarization tensor $\epsilon_{\mu \nu} $, with ${\epsilon^\mu}_\mu =0= k_1^\mu \epsilon_{\mu \nu}$,   whereas the outgoing graviton has momentum $k_2= (\omega_2 , \vec k_2 ) $ and polarization $\pi_{\mu \nu}$, with ${\pi^\mu}_\mu =0= k_2^\mu \pi_{\mu \nu}$. The amplitude $A_{(a)} +  A_{(b)}$, which is associated with  the diagrams $(a)$ and $(b)$ of Figure~2, is given by
\bea
A_{(a)} +  A_{(b)} =  && {\hskip - 5 truemm} 4 G \biggl \{  \frac{(p+k)^\mu (p+k)^\nu (p-k)^\sigma (p-k)^\tau}{2 ( p\cdot q + k^2) } \,  \pi^*_{\mu \nu} \epsilon_{\sigma \tau} \nonumber \\ 
&& \qquad -  \frac{(p-k)^\mu (p-k)^\nu (p+k)^\sigma (p+k)^\tau}{2 ( p\cdot q - k^2) } \,  \pi^*_{\mu \nu} \epsilon_{\sigma \tau}  \biggr \} \;  ; 
\label{3}
\eea
the amplitude $ A_{(c)}$ corresponding to the diagram $(c)$ is 
\bea
A_{(c)} = && {\hskip - 5 truemm} 8 G \biggl \{    p^\mu p^\nu \, \pi^*_{\mu \sigma} {\epsilon^\sigma}_\nu -  \left [ k^\mu k^\nu - \mezzo \eta^{\mu \nu} k^2\right ] \,  \pi^*_{\mu \sigma} {\epsilon^\sigma}_\nu \biggr \} \; ; 
\label{4}
\eea
finally the amplitude $ A_{(d)}$, which is represented by the diagram $(d)$,  reads 
\bea 
 A_{(d)} = && {\hskip - 5 truemm} 4 G \biggl \{  - \frac{( p\cdot q )^2}{4 k^2} \, \pi^*_{\mu \nu} \epsilon^{\mu \nu} + \left [ k^\mu k^\nu - \trequarti \eta^{\mu \nu} k^2\right ] \,  \pi^*_{\mu \sigma} {\epsilon^\sigma}_\nu - p^\mu p^\nu \, \pi^*_{\mu \sigma} {\epsilon^\sigma}_\nu  \nonumber  \\ 
 &&  \quad + 2 \frac{p^\mu p^\nu k^\sigma k^\tau}{k^2} \, \pi^*_{\mu \sigma} \epsilon_{\nu \tau}  -\frac{p^\mu p^\nu k^\sigma k^\tau}{k^2} \left [ \pi^*_{\mu \nu} \epsilon_{\sigma \tau} + \pi^*_{\sigma \tau} \epsilon_{\mu \nu} \right ]  \nonumber \\
&& \qquad + \frac{p\cdot q}{k^2} p^\nu k^\sigma \left [ \, \pi^*_{\nu \tau}{\epsilon^\tau}_\sigma - \epsilon_{\nu \tau} {{\pi^*}^\tau}_\sigma \right ]  \biggr \} \; , 
\label{5}
\eea
where the three independent momenta $p$, $q$ and $k$ are defined by 
\be
 p = \mezzo ( p_1 + p_2 ) \quad , \quad q = \mezzo ( k_1 + k_2 ) \quad , \quad 
 k = \mezzo ( p_1 - p_2) = \mezzo ( k_2 - k_1) \; , 
\label{6}
\ee
and the scattering angle $\theta $ is determined by $k^2 = - \omega_1 \omega_2 \sin^2 \theta / 2$. The individual terms $A_{(a)} $, $A_{(b)} $, $ A_{(c)}$ and $A_{(d)} $ are not gauge-invariant but their sum is. In the present context, gauge invariance means that $A_C$ vanishes when $\epsilon^{\mu \nu} $ is replaced by $k_1^\mu \xi^\nu + \xi^\mu k_1^\nu $ with arbitrary $\xi^\mu $ (and similarly for $\pi_{\mu \nu}$). To sum up, the Lorentz-invariant and gauge-invariant scattering amplitude $A_C$ turns out to be (see for instance \cite{GJ})
\bea 
A_C =&&  {\hskip - 5 truemm} 4 G \biggl \{  - \frac{( p\cdot q )^2}{4 k^2} \, \pi^*_{\mu \nu} \epsilon^{\mu \nu} - \left [ k^\mu k^\nu - \quarto \eta^{\mu \nu} k^2\right ] \,  \pi^*_{\mu \sigma} {\epsilon^\sigma}_\nu \nonumber  \\ 
&& {\hskip - 3 truemm} +  \frac{(p+k)^\mu (p+k)^\nu (p-k)^\sigma (p-k)^\tau}{2 ( p\cdot q + k^2) } \,  \pi^*_{\mu \nu} \epsilon_{\sigma \tau} + 2 \frac{p^\mu p^\nu k^\sigma k^\tau}{k^2} \, \pi^*_{\mu \sigma} \epsilon_{\nu \tau} \nonumber \\
&& -  \frac{(p-k)^\mu (p-k)^\nu (p+k)^\sigma (p+k)^\tau}{2 ( p\cdot q - k^2) } \,  \pi^*_{\mu \nu} \epsilon_{\sigma \tau} + p^\mu p^\nu \, \pi^*_{\mu \sigma} {\epsilon^\sigma}_\nu \nonumber \\
&& +\frac{p\cdot q}{k^2} p^\nu k^\sigma \left [ \, \pi^*_{\nu \tau}{\epsilon^\tau}_\sigma - \epsilon_{\nu \tau} {{\pi^*}^\tau}_\sigma \right ]  -\frac{p^\mu p^\nu k^\sigma k^\tau}{k^2} \left [ \pi^*_{\mu \nu} \epsilon_{\sigma \tau} + \pi^*_{\sigma \tau} \epsilon_{\mu \nu} \right ] \biggr \}
 \; . 
\label{7}
\eea

\subsection {Low-energy limit} 

We are interested in the case in which the graviton momenta  $k_1 $ and $k_2$  are small compared to the momentum of the massive particle, so that  in the semiclassical ---or large quantum numbers--- limit for each particle of dust, one can neglet  the variation of the momentum of the massive particle and one can put $p_1 \simeq p_2 \simeq p = E \, ( 1 , \vec v \, ) $, where $E$ represents  the energy of the particle and $\vec v$ denotes its  velocity.  We shall also concentrate on the coherent component of the gravitational scattering, in which the frequence of the outgoing gravitational wave is equal to the frequence of the incoming wave.  Then one has $ k_1 \simeq  \omega ( 1 , \widehat n_1 ) $ and $k_2 \simeq \omega ( 1 , \widehat n_2 ) $, where $\widehat n_1$ and $\widehat n_2$ are the unit vectors representing the directions along which the incoming and outgoing gravitational waves propagate. 

For fixed energy $E$ and fixed scattering angle  $\theta $, let us consider the low energy  $ (\omega / E ) \rightarrow 0 $ limit.  In the Taylor expansion of  the scattering amplitude (\ref{7}) in powers of $ (\omega / E ) $, the  leading term $B$  does not depend on $\omega $;  $B$ represents the effective amplitude which describes the low energy graviton scattering from a classical particle of dust. 
When the nontrivial components of the polarizations tensors $\pi_{\mu \nu }$ and $\epsilon_{\mu \nu}$ are of spatial type,  the expression of $B$,  up to second order in powers  of the velocity components $ \vec v $ of the massive particle,  is given by
\bea 
B (E,\vec v \, )= &&  {\hskip - 5 truemm}  \frac{G E^2}{ \sin^2 \theta/2} \biggl \{ 
\pi^*_{ij}\,  \epsilon^{ij} - \vec v \cdot ( \widehat n_1 + \widehat n_2 ) \, \pi^*_{ij} \, \epsilon^{ij} + 2 v^i n_2^j \, \pi^*_{ik}\,  {\epsilon_j}^k   + 2 v^i n_1^j  \, {\pi^*_j}^k \, \epsilon_{ik} 
 \nonumber  \\ 
&& {\hskip - 3 truemm} +  \mezzo \left [ (\vec v \cdot \widehat n_1 )^2 +  (\vec v \cdot \widehat n_2 )^2 \right ] \, \pi^*_{ij}\,  \epsilon^{ij} - 2 (\vec v \cdot \widehat n_2 ) v^i n_2^j\,  \pi^*_{ik} \, {\epsilon_j}^k 
 \nonumber \\
&& - 2 (\vec v \cdot \widehat n_1 ) v^i n_1^j \, {\pi^*_j}^k \, \epsilon_{ik} - 4 v^i v^j \, \pi^*_{ik} \, {\epsilon_j}^k \, \sin^2 \theta/2  + 2 v^i v^j n_1^k n_2^p\, \pi^*_{i k}\, \epsilon_{jp}
 \nonumber \\
&& +v^iv^j \left [  n_2^k n_2^p \, \pi^*_{ij} \, \epsilon_{kp} + n_1^k n_1^p \, \pi^*_{kp} \,  \epsilon_{ij} \right ] \biggr \}  \; . 
\label{8}
\eea
Latin indices $i,j, \cdots $ are used to denote the spatial components of the Lorentz vectors or tensors and take the values $i,j, \cdots =1,2,3$.  The sum over repeated latin indices is euclidean, i.e. $ a_j b^j = a^j b_j =  a_j b_j = a^j b^j = a_1 b_1 + a_2 b_2 + a_3 b_3 = \vec a \cdot \vec b $. 

In order to display the low energy  contribution of each Feynman diagram to the scattering amplitude,  let us denote by $B_{(\alpha )}$ the leading term of $A_{( \alpha )} $ in the $ (\omega / E ) \rightarrow 0 $ limit. One finds:
 \be
B_{(a)} +  B_{(b)} =   16 G E^2  \sin^2 \theta /2 \, \frac{v^i v^j v^k v^\ell }{\left [ 2 - \vec v \cdot ( \widehat n_1 + \widehat n_2 ) \right ]^2  }\pi^*_{ij} \epsilon_{k \ell } = O (v^4)  \; , 
\label{9}
\ee
\be
B_{(c)} = - 8 G E^2\,  v^i v^j \, \pi^*_{ik} \, {\epsilon_j}^k = O (v^2) \; , 
\label{10}
\ee
 \bea 
 B_{(d)} =  &&  {\hskip - 5 truemm}  \frac{G E^2}{ \sin^2 \theta/2} \biggl \{ 
\pi^*_{ij}\,  \epsilon^{ij} - \vec v \cdot ( \widehat n_1 + \widehat n_2 ) \, \pi^*_{ij} \, \epsilon^{ij} + 2 v^i n_2^j \, \pi^*_{ik}\,  {\epsilon_j}^k   + 2 v^i n_1^j  \, {\pi^*_j}^k \, \epsilon_{ik} 
 \nonumber  \\ 
&& {\hskip  3 truemm} +  \mezzo \left [ (\vec v \cdot \widehat n_1 )^2 +  (\vec v \cdot \widehat n_2 )^2 \right ] \, \pi^*_{ij}\,  \epsilon^{ij} - 2 (\vec v \cdot \widehat n_2 ) v^i n_2^j\,  \pi^*_{ik} \, {\epsilon_j}^k 
 \nonumber \\
&& - 2 (\vec v \cdot \widehat n_1 ) v^i n_1^j \, {\pi^*_j}^k \, \epsilon_{ik} + 4 v^i v^j \, \pi^*_{ik} \, {\epsilon_j}^k \, \sin^2 \theta/2  + 2 v^i v^j n_1^k n_2^p\, \pi^*_{i k}\, \epsilon_{jp}
 \nonumber \\
&& {\hskip 10 truemm} +v^iv^j \left [  n_2^k n_2^p \, \pi^*_{ij} \, \epsilon_{kp} + n_1^k n_1^p \, \pi^*_{kp} \,  \epsilon_{ij} \right ] \biggr \}  = \nonumber \\ 
= &&   {\hskip - 5 truemm}  \frac{G E^2} { \sin^2 \theta/2}\biggl \{   \pi^*_{ij}\,  \epsilon^{ij} 
- \vec v \cdot ( \widehat n_1 + \widehat n_2 ) \, \pi^*_{ij} \, \epsilon^{ij} \nonumber \\ 
&&  {\hskip 15 truemm } + 2 v^i n_2^j \, \pi^*_{ik}\,  {\epsilon_j}^k   + 2 v^i n_1^j  \, {\pi^*_j}^k \, \epsilon_{ik} \biggr \} + O (v^2) 
\; . 
\label{11}
\eea
The sum $B_{(a)} + B_{(b)} + B_{(c)} + B_{(d)}$,  up to second order in powers of the velocity components $\vec v$,  coincides with expression $B(E, \vec v) $ shown in equation (\ref{8}). 
 It is important to note that the amplitude contribution $B_{(a)} + B_{(b)}$, which is associated with the exchange (in the $s$- and $u$-channels) of the massive particle,  is  at least of fourth order in powers of the velocity. The contact term $B_{(c)}$ is quadratic in the velocity. 
The zero order term and first order term  ---in powers of the velocity--- originate from the graviton exchange in the $t$-channel exclusively, amplitude $B_{(d)}$. Finally, the $[ \sin^2\theta/2\, ]^{-1}$ factor in front of expression (\ref{11}), which  is due to the one-graviton exchange,  is strictly connected with the presence of the newtonian potential in gravitational interactions. 

\subsection{Many particles scattering}

In order to produce the amplitude $A$  for the graviton scattering from a many particles system, it is convenient to consider first,  in the case of a single-particle scattering, the wave packets representing the quantum mechanical states of the massive particle. The wave function of the initial state of the particle can be represented, for instance, by a gaussian wave packet that (at a fixed time) corresponds to an average momentum $\vec p_1 $ and an  average position $\vec x$, 
\be 
\psi_{\rm in} (\vec p \, ) \sim \, e^{-\alpha | \vec p - \vec p_1 |^2 } \; e^{- i \vec p \cdot \vec x } \quad . 
\label{12}
\ee
Similarly, the wave function of the final state is of the type 
\be 
\psi_{\rm out} (\vec p \, ) \sim \, e^{-\beta | \vec p - \vec p_2 |^2 } \; e^{- i \vec p \cdot \vec x } \quad . 
\label{13}
\ee
In the limit in which the gaussian wave functions become delta functions concentrated on the   momenta $\vec p_1 $ and $\vec p_2$, the two phase factors which are related to the position of the particle give origin to the following  contribution 
\be 
e^{i ( \vec p_2 - \vec p_1 ) \cdot \vec x } = e^{ i ( \vec k _1 - \vec k_2 ) \cdot \vec x } \; , 
\label{14}
\ee
which acts as a multiplicative factor on the  amplitude (\ref{8}). For a single-particle scattering, the presence of factor (\ref{14}) can be ignored, but in the case of a many particles scattering the position-dependent phase (\ref{14}) has to be taken into account.   

It should be noted that, for our puposes, the relativistic covariant generalization of the multiplicative phase factor (\ref{14}) is not needed. In fact, for the low energy coherent scattering of gravitons, the nonvanishing components of the momentum transfer are of spatial type, $ k_2 - k_1 \simeq \omega ( 0 , \widehat n_2 - \widehat n_1 )$.  The same conclusion can also be obtained by  means of the following argument. The leading order of the low-energy approximation in which $\omega_1 \simeq \omega_2$ is based on the assumption that, during the scattering process, the relevant global variables which are associated with the many particles system are essentially (or, can essentially be considered) constant in time; consequently, the scattering can be understood as a stationary process in which the graviton energy is  (in first approximation) conserved. In a stationary process, then, only the relative spatial positions of the different dust particles, as described by the phase factor ({\ref{14}), can appear in the expression of the amplitude. 

Let each particle of dust be labelled by the index $a$;  $E_a$ and $\vec v_a $ represent the energy and the velocity of the $a$-th particle. 
By taking into account the position-dependent phase factor (\ref{14}) and the normalization factor $E_a^{-1}$ which must  multiply  expression (\ref{8}), the resulting  amplitude for a many particles scattering is 
\be
A = \, \sum_a \frac{1}{E_a} \, B (E_a ,\vec v_a \, ) \, e^{i \, (\vec k_1 -\vec k_2 ) \cdot  \vec x_a } \; . 
\label{15}
\ee
 The coordinate system can always be chosen so that 
\be 
 \sum_a  E_a \, \vec v_a = 0 = \sum_a E_a \, \vec x_a \; , 
 \label{16}
 \ee
 and the graviton cross section takes the form
 \be
 \frac{d \sigma }{d \Omega} = |\, A \, |^2 = \left |  \sum_a \frac{1}{E_a} \, B (E_a ,\vec v_a \, ) \, e^{i \, (\vec k_1 -\vec k_2 ) \cdot  \vec x_a } \right |^2 \; . 
 \label{17}
 \ee 
 Equation (\ref{17}) gives the wanted expression of the cross section for the gravitons scattering from a composite classical matter system. 
 
 When the spatial extension $D$ of the system is bigger than  the wave length of the gravitational wave, $D > \lambda $, the exponential factor in (\ref{17}) has large fluctuations and the total cross section can be approximated by a sum of cross sections for the different parts of the system. On the other hand,  when $D < \lambda$, one can use the approximation $ e^{i \, (\vec k_1 -\vec k_2 ) \cdot  \vec x_a } \simeq 1 $.  In this case, one obtains 
\bea 
 A  = &&  {\hskip - 5 truemm}  \frac{G E_0}{ \sin^2 \theta/2} \biggl \{ 
\pi^*_{ij}\,  \epsilon^{ij} \left [ 1 + \mezzo W_{kp} (n_1^k n_1^p + n_2^k n_2^p ) + \Lambda ( 1 - 4 \sin^2\theta/2 ) \right ] 
 \nonumber  \\ 
&& {\hskip - 12 truemm} - 2 {W^i}_p \, n_2^p n_2^j\,  \pi^*_{ik} \, {\epsilon_j}^k -  2 {W^i}_p \,  n_1^p  n_1^j \, {\pi^*_j}^k \, \epsilon_{ik} +  2 W^{ij}\, n_1^k n_2^p\, \pi^*_{i k}\, \epsilon_{jp} + 2 \Lambda  n_1^i n_2^j\,  \pi^*_{ik} \, {\epsilon_j}^k 
 \nonumber \\
&& -  4 W^{ij} \, \pi^*_{ik} \, {\epsilon_j}^k \, \sin^2 \theta/2  
 + W^{ij} \left [  n_2^k n_2^p \, \pi^*_{ij} \, \epsilon_{kp} + n_1^k n_1^p \, \pi^*_{kp} \,  \epsilon_{ij} \right ] \biggr \}  \; , 
\label{18}
\eea
where 
\be
\sum_a E_a = E_0 \qquad , \qquad 
\sum_a E_a \, v^i v^j = E_0 \left ( W^{ij} + \Lambda \delta^{ij} \right ) \; , 
\label{19}
\ee
with $W^{ij} = W^{ji} $ and ${W_i}^i =0 $. 

\subsection{Corrections of first order in momentum transfer}
Let us now take into account the corrections to the scattering amplitude (\ref{18}) which are of  the first order in powers of the graviton momentum transfer;  that is, let us put $ e^{i \, (\vec k_1 -\vec k_2 ) \cdot  \vec x_a } \simeq 1 +  i \, (\vec k_1 -\vec k_2 ) \cdot  \vec x_a$ in equation (\ref{15}).  The part of the amplitude which is proportional to the momentum transfer gives the first correction to the geometric optics approximation. For slowly moving particles, one can neglect the terms with two or more powers of the velocity components $\vec v_a$, and from equation (\ref{15}) one finds  
\bea 
 A  = &&  {\hskip - 5 truemm}  \frac{G E_0}{ \sin^2 \theta/2} \biggl \{ 
\pi^*_{ij}\,  \epsilon^{ij} \left [ 1 + i \omega  \, \varepsilon_{k\ell p} \, S^k \, n_2^\ell \, n_1^p  \right ] 
 \nonumber  \\ 
&& {\hskip - 3 truemm} -i \omega \, \varepsilon_{\ell i p} \, S^p \, (n_2^\ell - n_1^\ell ) \left  [  n_2^j  \,  {\pi^*_k}^i \, {\epsilon_j}^k   + n_1^j \,  {\pi^*}_{jk} \, \epsilon^{ik} \right ]   
 \biggr \}  \; , 
\label{20}
\eea
where $\varepsilon_{ijk}$ denotes the completely antisymmetric 3-tensor and 
$J^i = E_0 S^i$ represents the angular momentum of the many particles system, 
\be
\sum_a E_a = E_0 \quad , \quad \sum_a E_a \, x_a^i \, v_a^j \; = \; \mezzo \varepsilon^{ijk}  J^k = \mezzo E_0 \, \varepsilon^{ijk} \, S_k \; . 
\label{21}
\ee
It should be noted that, in addition to the antisymmetric component (\ref{21}), the sum $\sum_a E_a x_a^i v_a^j $ could also contain a symmetric part $Q^{ij} = Q^{ji}$.  But since $Q^{ij}$ corresponds to a total time derivative, $Q^{ij} = (d / dt ) \mezzo \sum_a E_a x_a^i x_a^j $, it gives a vanishing contribution to the scattering amplitude  in the low-energy stationary approximation.   

When $\widehat n_1$ is directed as $\vec J$, the effect of the gravitational elicity interaction \cite{BE} is maximal. In fact, from equation (\ref{20}) it follows that the difference of the cross sections for the two different  helicities $\pm 2$ of the graviton becomes (in the $\theta \rightarrow 0 $ limit)
\be
 \frac{(d\sigma_{+}/d\Omega)-
(d\sigma_{-}/d\Omega)}{(d\sigma_{+}/d\Omega)
+(d\sigma_{-}/d\Omega)}\simeq -  \frac{2 \, \omega \, J\, \theta^2 }{E_0} \; . 
\label{22}
\ee

The important point now is that the amplitude (\ref{20}) only depends on the total energy $E_0$ of the dust system ---which is globally at rest--- and on its total angular momentum $J$. Therefore, expression (\ref{20})  should also represent the amplitude ---computed in the Born approximation--- for the scattering of gravitons from a rotating star or planet of mass $M=E_0$ and angular momentum $J$. In fact, equation (\ref{20}) is in agreement with the results which have been obtained,  by means of classical methods (see for instance  \cite{FHM, AJ, P, BE,  D1, D2}), for the transition amplitude associated with the gravitons scattering from a rotating star. As a check, let us consider the scattering of unpolarized gravitons  from a massive body whith $J=0$;  equations (\ref{17}) and (\ref20}) give 
\be
 \frac{d \sigma }{d \Omega} = \frac{ G^2 M^2}{\sin^4 \theta /2 } \left ( \cos^8 \theta/2 + \sin^8 \theta /2 \right )  \; , 
 \label{23}
 \ee 
which coincides precisely with Peters formula \cite{P}.

\section{Classical matter coupling}

By construction,  amplitude (\ref{20}) represents the semiclassical approximation  of a complete gauge-invariant transition amplitude for the  low energy scattering of gravitons. Expression (\ref{20}) is the sum of the terms of zero-order and first-order in powers of the velocity components of the dust particles. It has been shown in Section~2.2 that these two contributions originate from the one-graviton exchange diagram of Figure~2(d) exclusively. So one expects that, by taking the classical limit of the matter-gravity coupling in the lagrangian of effective quantum gravity,  the amplitude (\ref{20}) for the gravitational scattering of gravitons from a rotating massive body  could also be obtained by means of a single Feynman diagram containing a one-graviton exchange.

 This possibility has already been considered in the literature by De Logi and Kov\'acs \cite{K},  but the result produced in \cite{K} is not in agreement with equation (\ref{20}) and is not in agreement with Peters formula (\ref{23}).  As  admitted  in equation  (5.7)  of ref.\cite{K}, the cross section $(d\sigma / d \Omega )_{DLK} $ of De Logi and Kov\'acs  is related to the cross section $ (d\sigma / d \Omega )_{P} $ of Peters according to the equation $(d\sigma / d \Omega )_{DLK} =  \cos^2 \theta \,  (d\sigma / d \Omega )_{P}  $. 
 
We shall now discuss this subject and we shall do the calculation again;  firstly, the classical matter coupling in the effective quantum gravity lagrangian will be considered,  then the corresponding scattering amplitude will be computed. As a matter of facts,  it  turns out that the final result coincides with expression (\ref{20}) and is in complete agreement with Peters formula (\ref{23}), as it should be. 

In the large distance limit and to first order in $\ v/c  $, the coupling of the fluctuation field $  h_{\mu \nu } (x)  $ with a classical heavy rotating body, which is subject to stationary conditions and is placed in position $  \vec r  = \vec 0 $, is given ---at first order in $h_{\mu \nu}$--- by the action term 
\be
S_m  = - \mezzo \int d^4 x \, \Theta^{\mu \nu} (\vec x\, )  h_{\mu \nu} (x)
=  - \mezzo \int d^4 x  \left [ M  h_{0 0} (x) +  \epsilon^{ijk}  J_i
\partial_j h_{0k} (x)  \right ]  \delta^3 (\vec x  ) \, ,  
\label{24}
\ee
where $M  $ denotes the mass of the body and $ \{   J_i \}  $ are the components
of its total angular momentum. The coupling (\ref{24}) is in agreement with the expression of the metric which is induced by the presence of a rotating massive body \cite{LL, TEL}.  Since the energy-momentum tensor $\Theta^{\mu \nu} $ of the massive body which appears in equation (\ref{24}) is conserved, the interaction term $S_m$ is invariant under infinitesimal gauge transformations acting on $h_{\mu \nu }$. 

When the  tensor $\Theta^{\mu \nu} $  represents a fixed classical background field,  $\Theta^{\mu \nu} $ is no more a dynamical variable and then it does not transform under diffeomorphisms;  consequently,  the action $S_m$ together with the correction terms with higher orders in powers of $h_{\mu \nu}$ are not invariant under general coordinate transformations. This is just what happens in any non-abelian gauge theory in  the presence of a generic nontrivial background.  In order to clarify the connection between gauge-invariance and semiclassical limit, let us recall  the two possibilities: 

\noindent (a) one firstly computes the complete gauge-invariant transition amplitude,  and afterwards one takes the classical limit in the appropriate variables; 

\noindent (b)  one takes the appropriate classical limit directly in the interaction lagrangian, and subsequently one computes the transition amplitude. 

\noindent Method (a) always gives the correct answer for the quantities which are really observed in laboratories; this is precisely the way in which expression  (\ref{20}) has been derived in Section~2. Method (b) generally breaks gauge-invariance and leads to wrong conclusions, but there are exceptions. In fact, in this  Section it will be shown that, as far as the computation of the transition amplitude (at first order in $v/c$) for the gravitons scattering from a rotating star  is concerned, method (b) produces the correct answer (\ref{20}). As discussed in Section~2.2 and at the beginning of this section, this is essentially a consequence  of the low energy behavior of the amplitude components (\ref{9}), (\ref{10}) and (\ref{11}). 

The expansion of the lagrangian (\ref{1}) in powers of $h_{\mu \nu }$ determines the 3-gravitons vertex 
\bea
S^{(3)}_{EH}  &=&  {1\over 32 \pi G} \int  d^4x \,  \Bigl [ \partial_\mu h_\nu^{~ \nu} \partial^\mu h_{\sigma \tau} h^{\sigma \tau} + \quarto \partial_\mu h_{\sigma \tau} \partial^\mu h^{\sigma \tau} h_\nu^{~ \nu} \nonumber \\
&&  {\hskip - 0.4 truecm} -\mezzo \partial_\mu h_{\nu \sigma} \partial^\mu ( h^{\nu \tau} h_{\tau}^{~ \sigma})  - \quarto \partial_\mu h_\nu^{~ \nu} \partial^\mu h_{\sigma}^{~ \sigma} h_\tau^{ ~ \tau} - \partial_\mu h_\tau^{ ~ \tau} \partial^\nu ( h^{\sigma \mu} h_{\sigma \nu} ) \nonumber \\
&&  {\hskip - 0.4 truecm}  + \mezzo \partial_\mu h_\tau^{~ \tau} \partial_\nu ( h_\sigma^{~ \sigma} h^{\mu \nu} ) - \partial_\mu h_{\tau \sigma} h^{\tau \sigma} \partial_\nu h^{\mu \nu} + \partial_\mu h_{\tau \nu} \partial^\nu ( h^{\mu \sigma} h_\sigma^{~ \tau}) \nonumber  \\
&&  {\hskip - 0.4 truecm} + \partial_\mu h_{\tau \sigma} \partial_\nu h^{\mu \sigma} h^{\tau \nu} - \mezzo \partial_\mu h_{\tau \sigma} \partial_\nu h^{\tau \sigma}  h^{\mu \nu} - \mezzo \partial_\mu h_{\tau \nu} \partial^\nu h^{\mu \tau} h_\sigma^{ ~ \sigma} \Bigr ] \, . 
\label{25}
\eea 
With a covariant gauge-fixing,  the graviton propagator is  given by 
\bea
{\WT{h^{\mu \nu}(x)\> h}}_{\tau \sigma}(y)  = && {\hskip -0.5 truecm } i (16\pi G) \int {d^4p\over
(2\pi)^4}\, { e^{-ip(x-y)} \over p^2 + i  \epsilon} \Bigr \{  
\delta^\mu_\tau \, \delta^\nu_\sigma  +  \delta^\mu_\sigma \, \delta^\nu_\tau 
 -  \eta^{\mu \nu}  \eta_{\tau \sigma} \nonumber \\  &&  {\hskip -0.3 truecm } +  {\alpha -1\over 
p^2  +  i  \epsilon} \left [ \delta^\mu_\tau p^\nu p_\sigma + 
\delta^\mu_\sigma p^\nu p_\tau  +  \delta^\nu_\tau p^\mu p_\sigma + 
\delta^\nu_\sigma p^\mu p_\tau \right ]  \Bigr \} \, ,  
\label{26}
\eea
where $\alpha $ represents the gauge parameter; the choice $  \alpha =0  $ is the analogue of the Landau gauge in electrodynamics, whereas $ 
\alpha =1 $ is the analogue of the Feynman gauge. In our computations, $\alpha $ is left free.  With the classical matter coupling (\ref{24}), the amplitude $\cal A$ for the gravitons  scattering is determined by the one-graviton exchange diagram shown in Figure~3, 
\be
{\cal A}  = i^2 \, \langle k_2 , \pi |  \left (  S_{m}  [\, {\WT{ h \, ] \> S^{(3)}_{EH} [h}} , h, h ] \right )  | k_1 , \epsilon \rangle \, , 
\label{27}
\ee
and takes the form 
\bea
{\cal A}  =&& {\hskip -0.5 truecm } {i 2 GM  \omega_1 \, \delta (\omega_1 - \omega_2) \over  \pi | \vec k_1 - \vec k_2 \, |^2}
 \Bigl \{ \pi^*_{ij} \,  \epsilon^{ij}  \Bigl [ 1 + i(J^k / \omega_1 M) \,  \varepsilon_{k \ell m} \, k_2^\ell \, k_1^m \Bigr] \nonumber \\ 
&&   {\hskip -0.4 truecm } - i(J_n/\omega_1 M)(\vec k_2 -\vec k_1)_m \, \pi^*_{i \ell} \,  \epsilon^\ell_{j} \Bigl [ (\vec k_2 -\vec k_1)_j \varepsilon^{inm} - (\vec k_2 -\vec k_1)_i \varepsilon^{jnm} \Bigr ] \Bigr \} \, . 
\label{28}
\eea

\vskip 0.3 truecm

\begin{figure}[htbp]
\centerline{\includegraphics[width=1.30in]{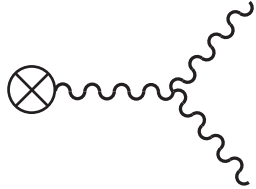}}
\caption{Feynman diagram corresponding to the low energy amplitude for gravitons scattering from a classical massive body.}
\label{F3}
\end{figure}

\vskip 0.3 truecm 

\noindent By means of the definition
\be
 {\cal A} = {i \, \delta  (\omega_1 - \omega_2) \over 2 \pi \omega_1 } \, A \, , 
\label{29}
\ee
the graviton scattering cross section is given by
\be
{d \sigma \over d \Omega } = | A  |^2 \, ,  
\label{30}
\ee
and, from expression (\ref{28}), one finds that the reduced amplitude $A$ is given by 
\bea
A  =&& {\hskip - 0.5 truecm}  { GM   \over  \sin^2(\theta /2 )}
 \Bigl \{ \pi^*_{ij}\, \epsilon^{ij}  \Bigl [ 1 + i(S^m / \omega_1 ) \varepsilon_{m j\ell} \, k_2^j\, k_1^\ell \Bigr]  \nonumber \\ 
&&   {\hskip -0.4 truecm } - i(S_n/\omega_1 )(\vec k_2 -\vec k_1)_m \, \pi^*_{i \ell} \,  \epsilon^\ell_{j} \Bigl [ (\vec k_2 -\vec k_1)_j \varepsilon^{inm} - (\vec k_2 -\vec k_1)_i \varepsilon^{jnm} \Bigr ] \Bigr \}  \, ,  
\label{31}
\eea
where $ S_k = J_k / M $ and $ | \vec k_2  - \vec k_1 \, |^2 = 4 \omega_1^2 \sin^2 (\theta / 2) $. For the scattering of gravitons from a classical rotating body, there are no additional contributions to the transition amplitude. In fact, the classical matter coupling at second order in powers of $h_{\mu \nu}$ takes the form $ \quarto \int d^4x \left [ (M/2) h_{00}^2(x) +\epsilon^{ijk} \, J_i  \partial_j (h_{00}(x)h_{0k}) \right ]  \delta^3 (\vec x  ) $ and gives a vanishing contribution for on-shell gravitons.  Similarly, there are no nontrivial corrections to the amplitude coming from the  gauge-fixing lagrangian. 

By taking into account the relations $\omega_1 = \omega_2 = \omega$, $\vec k_1 = \omega \, \widehat n_1 $ and $\vec k_2 = \omega \, \widehat n_2 $, expression (\ref{31}) can be written as 
\bea 
 A  = &&  {\hskip - 5 truemm}  \frac{G M}{ \sin^2 \theta/2} \biggl \{ 
\pi^*_{ij}\,  \epsilon^{ij} \left [ 1 + i \omega  \, \varepsilon_{k\ell p} \, S^k \, n_2^\ell \, n_1^p  \right ] 
 \nonumber  \\ 
&& {\hskip - 3 truemm} -i \omega \, \varepsilon_{\ell i p} \, S^p \, (n_2^\ell - n_1^\ell ) \left  [  n_2^j  \,  {\pi^*_k}^i \, {\epsilon_j}^k   + n_1^j \,  {\pi^*}_{jk} \, \epsilon^{ik} \right ]   
 \biggr \}  \; , 
\label{32}
\eea
which coincides with equation (\ref{20}) in which $ E_0 = M$. 

To sum up, because of the low energy dominance of the newtonian scattering,  the   transition amplitude (\ref{20}) for the gravitons scattering from a macroscopic rotating massive body can also be obtained  by means of the single Feynman diagram shown in Figure~\ref{F3}.  Therefore, the amplitude which corresponds to the Feynman diagram of Figure~3 is really in complete agreement with Peters formula (\ref{23}) and  represents the semiclassical approximation  of a gauge-invariant transition amplitude for the low energy scattering of gravitons.  

\section{Conclusions}

In this article, the cross-section for the scattering of gravitons from an extended macroscopic system made of classical massive bodies has been derived. By means of the effective quantum gravity formalism, the gauge-invariant scattering amplitude  has been computed in the Born approximation and the low energy limit has been discussed. 
With the inclusion of the corrections of the first order in powers of the graviton momentum transfer,  the transition amplitude of first order in the velocities only depends on the total energy and total angular momentum of the matter system and  coincides also with the amplitude for the graviton scattering from a rotating planet or star.  It has been shown that  the same result can also be obtained by considering,  in the effective quantum gravity approach, the classical limit for the lagrangian matter-gravitons  coupling. In this case, the transition amplitude corresponds to a single Feynman diagram with one 3-gravitons vertex and the result  coincides with the expression obtained in classical general relativity. 

\vspace{5mm} 

{\bf{Acknowledgments.}} I wish to thank Raymond Stora and the members of the Laboratoire d'Annecy-le-Vieux de Physique Th\'eorique  
for useful discussions.

\end{document}